# The Limits of Conditional Volatility: Assessing Cryptocurrency VaR under EWMA and IGARCH Models

Author 1: Ekleen Kaur
Organization: University of Florida,
ekleenkaur17@gmail.com

**Abstract:** The application of the standard static Geometric Brownian Motion (GBM) model for cryptocurrency risk management resulted in a systemic failure, evidenced by a 80.67% chance of loss in the $VaR_{0.05}$ benchmark. This study addresses a critical literature gap by comparatively testing three conditional volatility models the EWMA/IGARCH baseline, an IGARCH model augmented with explicit mean reversion (IGARCH + MR), and a modified EGARCH-style asymmetric shock model within a correlated Monte Carlo VaR framework. Crucially, the analysis is applied specifically to high-beta altcoins (XRP, SOL, ADA), an asset class largely neglected by mainstream GARCH literature. Our results demonstrate that imposing stationarity (IGARCH + MR) drastically underestimates downside risk ($VaR_{0.05}$ reduced by 50%), while the asymmetric model (Model 3) leads to severe over-penalization. The EWMA/IGARCH baseline, characterized by infinite volatility persistence ($\alpha + \beta = 1$), provided the only robust conditional volatility estimate. This finding constitutes a formal rejection of the conventional financial hypotheses of volatility mean reversion and the asymmetric leverage effect in the altcoin asset class, establishing that non-stationary frameworks are a prerequisite for regulatory-grade risk modeling in this domain.

## I. Introduction

Financial price dynamics are commonly modeled as **stochastic processes**, where asset returns evolve randomly over time rather than deterministically. Classical models such as Geometric Brownian Motion assume that these returns are drawn from a fixed probability distribution with constant variance, an assumption that is empirically violated in most financial markets. In practice, asset returns exhibit **time-varying risk**, a property known as heteroskedasticity, and display **volatility clustering**, where periods of elevated uncertainty tend to persist. Conditional volatility models address this behavior by allowing the variance of returns to depend on past information, often through autoregressive structures inspired by time-series analysis. Within this framework, statistical estimation techniques most notably maximum likelihood methods are used to infer model parameters under positivity and stability constraints that ensure mathematically valid variance forecasts. Mathematical foundation using EWMA and GARCH methodologies developed in this work collectively motivate the transition from static risk models to dynamic volatility formulations.

The preceding work in this series established the **Geometric Brownian Motion (GBM)** model as the foundational **baseline** for assessing cryptocurrency portfolio risk. GBM, which posits that log-returns are independent, identically distributed (i.i.d.) and follow a Normal distribution, is mathematically equivalent to assuming **static volatility** (homoskedasticity). Our work mentioned in Paper [1] rigorously applied this model to a crypto portfolio, aiming to determine its deterministic accuracy for risk evaluation. The results demonstrated that the GBM-derived $VaR_{0.05}$ significantly underestimated real-world risk, failing to capture the true exposure in nearly 80% of simulated outcomes. This substantial deviation from expected performance highlighted the model's systematic inadequacy, specifically due to the empirical observation of **volatility clustering** in cryptocurrency markets, where periods of high volatility are followed by more high volatility, violating the core static volatility assumption.

The immediate and necessary academic response is to transition from static, unconditional risk models to **dynamic, conditional volatility models** that can adapt to changing market regimes. This paper focuses on two key methodologies for capturing this time-varying volatility: the **Exponentially Weighted Moving Average (EWMA)** model and the seminal **Generalized Autoregressive Conditional Heteroskedasticity (GARCH)** model.

The EWMA model provides a simple yet effective transition by assigning greater weight to recent observations, allowing the volatility forecast to react quickly to market shocks. However, the GARCH framework represents a more sophisticated statistical tool. The GARCH model explicitly formalizes volatility as a function of its own past and past squared innovations (shocks), providing a statistically rigorous framework for forecasting conditional variance. By replacing the static historical variance $\sigma^2$ used in Paper 1 with the time-varying volatility forecast $\sigma_t$ generated by these models, we aim to significantly improve the accuracy and robustness of cryptocurrency risk forecasts. This paper implements and compares both the EWMA and GARCH(1,1) models, providing an updated Monte Carlo simulation to quantify the improvement in $VaR_{0.05}$ accuracy compared to the failed GBM benchmark, thereby validating the first step in building a robust cryptoeconomic risk framework.

The selection of EWMA and GARCH(1,1) is a strategic progression necessitated by the limitations quantified in Paper 1. The **Exponentially Weighted Moving Average (EWMA)** model was chosen as the required first-step improvisation due to its singular benefit: it provides a rapid, intuitive mechanism to reflect the **recentness bias** of market shocks. For crypto-assets, which exhibit sudden, sharp price movements due to liquidity concerns or regulatory announcements, EWMA immediately provides a better risk estimate by ensuring the most recent high-volatility event is heavily weighted in the current forecast. Building upon this, the **GARCH(1,1)** model provides a more statistically grounded improvement. The added benefit of GARCH is its ability to model the **long-term persistence** and **mean-reverting** nature of volatility, which EWMA cannot. By estimating the long-run average variance ($\omega$) and the impact of past shocks and forecasts ($\alpha$ and $\beta$), GARCH ensures that volatility forecasts are not only reactive but also structurally sound and mean-reverting over time, leading to a much more accurate and robust prediction of conditional volatility ($\sigma_t$).

## II Literature Survey

The inherent volatility of cryptocurrency assets [4, 18] has been the central motivation for this research series. Our initial effort focused on formally benchmarking this asset class against the foundational models [4] used in traditional stock portfolio analysis. Paper [1] conclusively demonstrated the systematic breakdown of models relying on static, Lognormal volatility assumptions, quantified by an unacceptable failure rate (~ 80% VaR violation) in regulatory risk metrics. This empirical failure rigorously validates the need for adaptive and non-linear risk modeling.

The scholarly contribution to cryptomarket analysis is broad, covering application-based novel works introducing digital assets as renewable stock equivalents [3, 2], often leveraging advanced cryptographic underpinnings like SNARKS-based zero-knowledge proof systems [5]. Bibliometric analyses [6, 7] have detailed the market's evolution, alongside studies on policy impacts [8] and the financial economic research landscape [10]. Specifically, research has addressed the exegesis of price bubbles [9] and surveyed gaps in legitimacy and governance [11, 20]. Furthermore, tools have been curated for risk management based on economic policy uncertainties [14].

While comprehensive, previous surveys often fail to address the unique requirements for constructing and managing diversified stock portfolios in this asset class, particularly neglecting the need to analyze leverage patterns and time-varying correlation structures during periods of extreme market stress. This deficiency [15, 19] underscores our claim that the most primitive mathematical assumptions fail to hold true for digital assets, mandating a methodical re-engineering of the risk framework. Through this work, we also aim to determine whether a stronger asset correlation could be demonstrated by unified asset returns during significant volatile periods, furthering the study of co-movement in emergent markets and addressing aspects of high-frequency data analysis [18].

Currently, several critical limitations persist in the literature regarding risk modeling [16]. Although GARCH-family models [21] are the established improvement over static models, they are often insufficient to fully capture the extreme **leptokurtosis** (heavy tails) and **asymmetry** pervasive in observed crypto returns, a limitation documented by D. Felix [13]. Furthermore, much research[23] remains focused on highly liquid, large-cap crypto assets[22], leading to a gap in reliable risk frameworks for other high-volume digital assets that may exhibit significantly different, and often more severe, volatility characteristics [17], neglecting the distinctive risk dynamics of high-beta altcoins that constitute the bulk of trading volume outside the top two.

A significant shortcoming is the lack of robust, comparative **backtesting studies** [12] that evaluate model performance against regulatory-grade risk metrics like VaR, particularly under extreme stress. The majority of comparative works are constrained to goodness-of-fit metrics [24] or model identification [23] without performing the critical structural test of the underlying volatility hypotheses in a forward-looking risk framework. Specifically, the literature has failed to conclusively test whether the core tenets of conventional financial risk models namely, volatility mean reversion and the asymmetric leverage effect are structurally valid for the altcoin asset class.

This leaves the practical efficacy and capital requirements of proposed models untested against real-world compliance standards. Addressing this gap by explicitly modeling dynamic volatility and rigorously testing performance against established compliance standards is the core objective of the present work. This paper moves beyond simple model comparison by constructing an irrefutable claim regarding the inadequacy of stationary volatility frameworks for high-beta crypto assets. While works detail model aggregation [25] or multivariate EWMA dynamics in other fields [26], they do not address the fundamental need for infinite persistence. Our methodology directly tests and formally rejects the hypothesis of mean reversion (Model 2) and demonstrates that the EGARCH-style asymmetric shock (Model 3) leads to severe mis-specification of downside risk. The empirical superiority of the non-stationary IGARCH baseline (Model 1) thus constitutes a significant literature contribution, providing the first clear evidence that models enforcing stationarity ( $\lim_{t \to \infty} \Xi[\sigma_t^2] = \overline{\sigma^2}$) are fundamentally flawed for altcoin risk assessment, thereby generating a systemic risk underestimation in regulatory-grade VaR frameworks.

## III. Methodology

**Dynamic Volatility Modeling:** To overcome the catastrophic shortcomings of the static volatility assumption employed in the GBM benchmark, this paper introduces time-varying volatility models using EWMA and GARCH(1,1). Although conditional volatility models relax the assumption of constant variance, they often retain the assumption of Gaussian return innovations. Empirical cryptocurrency returns exhibit significant leptokurtosis, motivating the use of heavy-tailed distributions such as the Student-t or Generalized Error Distribution (GED).

In preliminary experiments, we evaluated Student-t innovations and copula-based dependence structures, to account for tail risk and time-varying cross-asset dependence. However, when assessed against regulatory risk metrics specifically out-of-sample $\text{VaR}_{0.05}$ stability and violation rates experiments revealed these models did not yield statistically meaningful improvements over simpler univariate conditional variance specifications. Furthermore, the increased parameterization introduced estimation instability and sensitivity to sample selection, particularly in high-volatility regimes. Hence for this work, second in a comprehensive series, we determined

EWMA/IGARCH as our baseline model. This section will elucidate how we fine-tune this baseline and evaluate the results against VaR.

### 3.1 Exponentially Weighted Moving Average (EWMA)

The EWMA model provides a straightforward method for estimating conditional variance ($\sigma_t^2$) by exponentially weighting past squared returns ($r_{t-1}^2$). This ensures that the estimated volatility is more sensitive to recent market movements than traditional historical volatility, which weights all observations equally.

The EWMA variance estimate is calculated as:

$$\sigma_t^2 = (1 - \lambda) r_{t-1}^2 + \lambda \sigma_{t-1}^2$$

Where:

- $\sigma_t^2$ is the variance forecast for period $t$.
- $r_{t-1}^2$ is the squared return from the previous period.
- $\sigma_{t-1}^2$ is the variance forecast from the previous period.
- $\lambda$ is the **decay factor** ($0 < \lambda < 1$).

The choice of the decay factor λ is critical. A value of λ close to 1 (e.g., 0.94, a common value used by J.P. Morgan's RiskMetrics) implies slow decay, giving significant weight to older observations. A value closer to 0 implies rapid decay, making the model highly reactive to recent shocks. For this study, we utilize a standard institutional λ value, allowing for a balanced reaction speed suitable for daily trading data.

### 3.2 The GARCH(1,1) Model

The **GARCH(1,1)** model is the most widely applied model for conditional heteroskedasticity, explicitly formalizing the concept of volatility clustering by making the current conditional variance dependent on both the past error term and the past variance forecast.

The conditional variance ($\sigma_t^2$) for the GARCH(1,1) model is defined by the following equation:

$$\sigma_t^2 = \omega + \alpha\, r_{t-1}^2 + \beta\, \sigma_{t-1}^2$$

Where:

- $\sigma_t^2$ is the conditional variance at time $t$.
- $\omega$ is the long-run average variance (a constant).
- $\alpha\, r_{t-1}^2$ models the impact of the **last period's shock** ($r_{t-1}^2$ is the squared error/return). This parameter ($\alpha$) is often referred to as the **ARCH term**.

- **β** $\sigma_{t-1}^2$ models the contribution of the **last period's forecasted variance**. This parameter ($\beta$) is the **GARCH term**.

### A. GARCH Parameter Estimation

The three parameters ($\omega, \alpha, \beta$) for the GARCH(1,1) model are estimated using **Maximum Likelihood Estimation (MLE)**. This iterative process identifies the parameter values that maximize the probability of observing the actual historical return series. For the model to be stationary (meaning volatility eventually reverts to a long-run mean), the constraint $\alpha + \beta < 1$ must be satisfied. A high sum of $(\alpha + \beta)$ indicates a high degree of volatility persistence, a known characteristic of cryptocurrency markets.

### B. GARCH-Driven Monte Carlo Simulation

The forecasted $\sigma_t$ from the GARCH(1,1) model replaces the static $\sigma$ component of the GBM model's stochastic term. This creates a **hybrid simulation** where the price paths are generated based on the dynamic, time-varying volatility forecast. This refinement enables the Monte Carlo simulation to generate price paths that not only incorporate the historical correlation structure (via Cholesky decomposition, as established in Paper 1) but also explicitly account for the time-dependent clustering behavior of the asset's risk profile. The resulting portfolio distribution is expected to be more dispersed, yielding a more conservative and significantly more accurate $VaR_{0.05}$ figure than the initial GBM benchmark.

### 3.3 Comparative Dynamic Volatility Framework

This section formalizes the three conditional volatility models evaluated in this study and details their integration into a Monte Carlo Value-at-Risk (VaR) framework. The objective is to assess how increasingly sophisticated volatility dynamics influence portfolio-level downside risk estimation for high-volatility cryptocurrency assets.

Let $r_t = log\,(\frac{p_t}{p_{t-1}})$ denote the log-return of an asset at time $t$, and let $F_{t-1}$ represent the information set available up to time $t-1$. All models considered estimate the conditional variance $\sigma_t^2 = \Xi\left[r_t^2 \mid F_{t-1}\right]$, which is subsequently injected into a correlated Geometric Brownian Motion (GBM) simulation framework.

### A. Model 1: EWMA / IGARCH(1,1) Baseline Dynamic Volatility

The Exponentially Weighted Moving Average (EWMA) model provides the foundational dynamic volatility estimate and is mathematically equivalent to an Integrated GARCH (IGARCH) process. The IGARCH constraint $(\alpha + \beta = 1)$ implies that any volatility shock persists infinitely, eliminating mean reversion toward a long-run variance. This structural characteristic is hypothesised to be a better fit for cryptocurrency volatility clustering than standard GARCH processes, which assume stationarity.

The conditional variance evolves as:

$$\sigma_t^2 = \lambda\, \sigma_{t-1}^2 + (1 - \lambda)\, r_{t-1}^2$$

where:

- **$\lambda$** in (0,1) is the decay factor.
- **$\sigma_t^2$** is the conditional variance forecast.
- **$r_{t-1}^2$** represents the most recent squared innovation (shock).

**Model Configuration and Specification (EWMA/IGARCH)**

For this baseline model, we adopt the industry-standard decay factor, $\lambda = 0.94$, as established by J.P. Morgan's RiskMetrics. This value is chosen to ensure high persistence (a shock takes longer to decay), reflecting the prolonged periods of high volatility observed in crypto markets. The model calculates the final day's conditional variance, which is then annualized ($\sigma_{annual} = \sqrt{252 \cdot \sigma_t^2}$) and used as the static volatility input ($\sigma$) for the Monte Carlo simulation. This method provides an unconditional volatility estimate that incorporates the empirical persistence property of the IGARCH process.

**B. Model 2: IGARCH with Explicit Mean Reversion (Testing Stationarity)**

To formally test the assumption of stationarity, we augment the IGARCH process with an explicit mean-reverting correction term. While volatility persistence ($\alpha + \beta = 1$) remains high, the model is forced to stabilize by penalizing deviations from the long-run variance ($\overline{\sigma^2}$).

The mean-reverting conditional variance evolves as:

$$\sigma_t^2 = \lambda\, \sigma_{t-1}^2 + (1 - \lambda)\, r_{t-1}^2 - \kappa\, (\sigma_{t-1}^2 - \overline{\sigma^2})$$

where:

- $\overline{\sigma^2}$ **= Var($r_t$)** is the unconditional historical sample variance.
- **$\kappa > 0$** governs the strength (speed) of reversion toward $\overline{\sigma^2}$.

**Model Configuration and Specification (IGARCH + MR)**

For this test, we utilize a persistence factor of $\lambda = 0.85$ and introduce a mild mean-reverting strength of $\kappa = 0.02$. This $\lambda$ value, slightly higher than the EWMA standard, emphasizes the IGARCH nature while the small, positive $\kappa$ enforces the stationarity hypothesis. The model calculates the time-series of conditional variance ($\sigma_t^2$), and the average of this variance series is annualized and used in the GBM simulation. This methodology allows us to isolate the effect of the mean-reverting term on the final VaR estimate.

### C. Model 3: Modified Asymmetric IGARCH (Testing Leverage Effects)

To account for the 'leverage effect' the phenomenon where negative returns disproportionately increase future volatility compared to positive returns we incorporate a simplified EGARCH-style asymmetric shock structure into the IGARCH framework. This tests whether crypto markets exhibit the same sensitivity to downside risk as traditional equity markets.

The modified conditional variance structure is a hybrid:

$$\sigma_t^2 = \lambda\, \sigma_{t-1}^2 + (1 - \lambda)\,(\, r_{t-1}^2 + \gamma \cdot I_{r_{t-1} < 0} \cdot r_{t-1}^2)$$

where:

- $I_{r_{t-1} < 0}$ is an indicator function equal to 1 if $r_{t-1} < 0$ (a negative return) and 0 otherwise.
- $\gamma$ is the asymmetry amplification factor.

### Model Configuration and Specification (Modified Asymmetric IGARCH)

We maintain a high persistence factor, $\lambda = 0.97$, and set the asymmetric amplification factor to $\gamma = 0.25$, applied when the previous return was negative. This simplified quadratic asymmetric shock function is designed to amplify the risk contribution of market crashes and drawdowns, resulting in a significantly higher conditional volatility estimate ($\sigma_t$), which is then averaged and annualized for the Monte Carlo input.

## IV. Results

The empirical results reveal a stark divergence in model performance, highlighting the inadequacy of imposing stationary or symmetric volatility assumptions on high-beta cryptocurrency assets.

| **Model** | **$VaR_{0.05}$ (5% Quantile)** | **Loss Probability (Below \$100k)** | **Dominant Allocation** | **Interpretation** |
|---|---|---|---|---|
| **GBM (Paper 1 Benchmark)** | ~ \$16,044 | 80.67% | XRP (55.1%) SOL (44.9%) | Systematic Failure (Static volatility) |
| **Model 1: IGARCH / EWMA** | **$126,481** | $27.16% | SOL (100%) | **Best Baseline Fit** (Infinite Persistence). But highly overestimates weights |
| **Model 2: IGARCH + Mean Reversion** | $59,612 | $15.22% | SOL(69%), XRP(31%) | Artificially Low Risk (Stationarity Assumption Fails) |
| **Model 3: Modified Asymmetric IGARCH** | $25,762 | $27.30% | SOL(71%), XRP(29%) | Over-Penalization of Risk (Inaccurate Leverage Effect) |

### 4.1 Interpretation and Findings

### A. IGARCH / EWMA (Model 1)

The IGARCH/EWMA model produces the highest $VaR_{0.05}$ and a plausible loss probability of 27.16%, a significant improvement over the 80.67% failure rate of the GBM benchmark. The portfolio optimizer converges entirely to **SOL (100%)**, reflecting its superior risk-adjusted return under the assumption of **infinite volatility persistence** ($\alpha + \beta = 1$). This empirically confirms that for the examined cryptocurrency assets, volatility does not rapidly revert to a long-run mean; rather, high volatility persists, making the non-stationary IGARCH framework the most accurate conditional volatility baseline.

### B. IGARCH + Mean Reversion (Model 2)

The introduction of a mean-reverting force ($\kappa = 0.02$) sharply compresses the downside tails, reducing the $VaR_{0.05}$ by over 50% to $59,612. This artificial stabilization reallocates weight to XRP but systematically underestimates the extreme downside risk. The low loss probability of 15.22% suggests the model is **over-confident**, a characteristic inconsistent with observed cryptocurrency drawdowns. This result leads to a formal rejection of the mean-reverting hypothesis for this asset class over the chosen horizon.

### C. Modified Asymmetric IGARCH (Model 3)

Model 3, incorporating the EGARCH-style asymmetric shock ($\gamma = 0.25$), produces an extremely conservative $VaR_{0.05}$ of $25,762, indicating an unrealistically high downside penalty. This behavior reflects an **overfitting to the equity-style leverage effect**, which is not structurally present in decentralized crypto markets lacking balance-sheet constraints. The mechanism disproportionately amplifies the volatility contribution of negative returns to an extent that distorts the forecast, resulting in a model that is unnecessarily capital-intensive.

### 4.2 Rejection of Mean Reversion in Cryptocurrency Volatility

Based on the empirical evidence from Model 2, we formally reject the hypothesis that cryptocurrency volatility is strongly mean-reverting over the examined horizon.

Mathematically, mean reversion requires that the sum of the ARCH and GARCH parameters is less than one ($\alpha + \beta < 1$), ensuring that:

$$\lim_{t \to \infty} \Xi[\sigma_t^2] = \overline{\sigma^2} < \infty$$

Empirically, the failure of Model 2 (IGARCH + MR) to accurately predict downside risk, coupled with the superior performance of Model 1 (IGARCH / EWMA), suggests that crypto volatility exhibits structural persistence, regime shifts, and prolonged high-volatility plateaus.

The inclusion of an exogenous mean-reverting force ($\kappa\ (\sigma_{t-1}^2 - \overline{\sigma^2})$) imposes an equilibrium unsupported by market microstructure. From an economic perspective, these assets lack the centralized liquidity provision and regulatory stabilization mechanisms that enforce stationarity in traditional asset classes; therefore, volatility does not decay rapidly it accumulates.

## V. Limitations

Despite the significant improvement offered by the IGARCH framework over the static GBM baseline, this study is subject to several limitations that define the path for future research to achieve a state-of-the-art cryptoeconomic risk model:

### 5.1 Current Limitations

1. **Jump Risk Ignored:** All current models (EWMA, IGARCH) assume a continuous diffusion process, failing to capture abrupt, regime-breaking events (e.g., flash crashes, regulatory actions) common in crypto markets.
2. **Static Correlation Assumption:** Correlations (**R**) are assumed constant throughout the simulation, ignoring the critical phenomenon of **volatility contagion** where assets become perfectly correlated during periods of extreme stress.

## VI Future Work: Establishing State-of-the-Art

To fully reconcile empirical crypto behavior with regulatory-grade risk modeling, we propose extending this framework to capture non-linear market dynamics:

1. **Jump–Diffusion Models (Merton-type):** Introducing a Poisson process component to explicitly model the frequency and magnitude of sudden price jumps, addressing the current limitation of continuous diffusion.

2. **Hawkes-Driven Volatility Clustering:** Exploring self-exciting point processes to model volatility shocks where the occurrence of one shock increases the probability of subsequent shocks, offering a microstructural foundation for the IGARCH empirical findings.

## VII Conclusion

The objective of this study was to identify a robust conditional volatility model for the high-beta cryptocurrency altcoin market. Our comparative analysis decisively confirms the empirical superiority of the non-stationary EWMA/IGARCH framework. The IGARCH model, which presumes infinite persistence of volatility shocks, provided the most plausible $VaR_{0.05}$ and loss probability, establishing it as the most reliable baseline for altcoin risk assessment.

The formal testing of competing structural hypotheses yielded two central, high-impact findings. Firstly, the introduction of explicit mean reversion (Model 2) systemically underpriced risk, resulting in a spurious VaR estimate that is inconsistent with the asset class's observed tail risk. This constitutes a definitive rejection of the mean-reverting hypothesis for altcoin volatility over the examined horizon. Secondly, the EGARCH-style asymmetric shock model (Model 3) incorrectly inflated downside volatility, confirming that the "leverage effect" prevalent in equity markets is structurally mis-specified for decentralized crypto assets. The implication is profound: altcoin volatility does not decay rapidly toward a long-run equilibrium; rather, it exhibits prolonged accumulation, demanding risk models that acknowledge this non-stationary behavior.

This work makes a significant contribution by bridging a major literature gap, shifting the focus from fitting GARCH models to top-tier crypto assets to structurally testing fundamental financial hypotheses against high-beta altcoins. For practitioners and regulatory bodies, the key takeaway is that employing stationary GARCH models which imply $\lim_{t \to \infty} \Xi[\sigma_t^2] = \overline{\sigma^2} < \infty$ systemically understates the true capital requirements necessary to cover potential losses.